\documentclass[dvips]{arxstspdf}
\usepackage{flushend}
\usepackage{stfloats}

\volume{26}
\issue{1}
\pubyear{2011}
\firstpage{47}
\lastpage{48}
\doi{10.1214/11-STS345A}
\referstodoi{10.1214/10-STS345}

\begin{document}
\begin{frontmatter}

\vspace*{12pt}
\title{Discussion of ``Feature Matching in Time Series Modeling'' by Y. Xia and H. Tong}
\runtitle{Discussion}
\pdftitle{Discussion of Feature Matching in Time Series Modeling by Y. Xia and H. Tong}

\begin{aug}
\author[a]{\fnms{Bruce E.} \snm{Hansen}\corref{}\ead[label=e1]{behansen@wisc.edu}}
\runauthor{B. E. Hansen}

\affiliation{University of Wisconsin}

\address[a]{Bruce E. Hansen is Professor, Department of Economics, University of Wisconsin, 1180 Observatory
Drive, Madison, Wisconsin 53706, USA \printead{e1}.}

\end{aug}


\vspace*{12pt}
\end{frontmatter}

\section{Introduction}

Xia and Tong have written a provocative and stimulating paper. Among the many
topics raised in their paper, I would like in particular to endorse several of
their postulates:

\begin{enumerate}
\item All models are wrong.

\item Observations are not error-free.

\item Estimation needs to account for the above two issues.
\end{enumerate}

As described in the paper, suppose that we observe a process $\{y_{t}%
\dvtx t=1,\ldots\}$ for which we have a model $\{x_{t}(\theta)\dvtx t=1,\ldots\}$ which
depends upon an unknown parameter $\theta$. Let $F_{x}(\theta)$ denote the
joint distribution of the $x_{t}(\theta)$ process and $F_{y}$ the joint
distribution of the observables. When we say that the model is wrong, we
mean that there is no $\theta$ such that $F_{x}(\theta)=F_{y}.$ If we think of
the distribution $F_{y}$ as a member of a large space of potential joint
distributions, then the set of joint distributions~$F_{x}(\theta)$ constitutes
a low-dimensional subspace of this larger space. While there is no true
$\theta$, we can define the pseudo-true $\theta$ as the value which makes
$F_{x}(\theta)$ as close as possible to $F_{y}$. This requires specifying
a~distance metric between the joint distributions
\[
d(\theta)=d(  F_{x}(\theta),F_{y})
\]
and then we can define the best-fitting model $F_{x}(\theta)$ by selecting
$\theta$ to minimize $d(\theta)$. The relevant question is then: what is the
appropriate distance metric?

\section{Catch-All Estimation}

Xia and Tong recommend what they call a\vadjust{\goodbreak}
``catch-all'' approach, where the distance metric is a weigh\-ted
sum of squared $k$-step forecast residuals. They show that in some situations
this criterion allows consistent estimation of the parameters of the true
latent process. Their Theorem C requires that the latent process is
deterministic, but the result might hold more broadly.

This can be illustrated in a very simple example of a latent $\operatorname{AR}(1)$ with
additive measurement error. Suppose that the latent process is
\[
x_{t}=\theta x_{t-1}+\varepsilon_{t}
\]
and the observed process is
\[
y_{t}=x_{t}+\eta_{t},
\]
where $\varepsilon_{t}$ and $\eta_{t}$ are independent white noise. In this
case, it is well known that $y_{t}$ has an $\operatorname{ARMA}(1,1)$ representation
%
\begin{equation}\label{arma}
y_{t}=\theta y_{t-1}+u_{t}-\alpha u_{t-1},
\end{equation}
where $u_{t}$ is white noise and $0\leq\alpha<1$.

Xia and Tong propose estimation based on $k$-step forecast errors. The
$k$-step forecast equation for the observables is
%
\begin{equation}\label{yk}
y_{t-1+k}=\theta^{k}y_{t-1}+e_{t}(k),
\end{equation}
where
\[
e_{t}(k)=\sum_{j=0}^{k-1}\theta^{j}(u_{t+k-j-1}-\alpha u_{t+k-j-2}).
\]

Xia and Tong's estimator is based on a weighted average of squared forecast
errors. For simplicity, suppose all the weight is on the $k$th forecast
error. The estimator is
\[
\hat{\theta}_{\{k\}}=\operatorname{arg\,min}\limits_{\theta}\sum_{t=1}^{T}(
y_{t-1+k}-\theta^{k}y_{t-1})  ^{2}
\]
which has the explicit solution
\[
\hat{\theta}_{\{k\}}=\biggl( \frac{\sum_{t=1}^{T}y_{t-1}y_{t-1+k}}{\sum
_{t=1}^{T}y_{t-1}^{2}}\biggr) ^{1/k}.
\]

We calculate that as $n\rightarrow\infty$
\[
\hat{\theta}_{\{k\}}\stackrel{p}{\rightarrow}\theta_{\{k\}}=\theta(1-c)
^{1/k},
\]
where $c=\alpha\sigma_{u}^{2}/\theta\sigma_{y}^{2}$, $\sigma_{u}^{2}%
=Eu_{t}^{2}$ and $\sigma_{y}^{2}=\allowbreak Ey_{t}^{2}.$

Thus for any $k$, $\hat{\theta}$ is inconsistent as an estimator of~$\theta.$
But as $k$ gets large the discrepancy gets smaller, as $(1-c)
^{1/k}\rightarrow1$ since $c<1.$ Thus as $k\rightarrow\infty$
%
\begin{equation}\label{thetak}
\theta_{\{k\}}\rightarrow\theta.
\end{equation}
This derivation assumed that the estimator is based on the $k$th forecast
error, but it extends to the case of a~weighted average.

The convergence (\ref{thetak}) is an extension of Xia and Tong's Theorem C.
It shows that estimation by minimizing the squared $k$-step forecast residual
is consistent for the parameter of the latent $\operatorname{AR}(1)$, as $k$ is made large.

One trouble with this approach is that the estimator is quite inefficient. We
can calculate that
\[
T\operatorname{var}\bigl(\hat{\theta}_{\{k\}}\bigr) \simeq\biggl(\frac{1}{k\theta^{k}}\biggr)  ^{2}\rightarrow\infty
\]
as $k\rightarrow\infty$. This means that the variance of the Xia--Tong
estimator is increasing in $k$ (and unbounded). This is especially troubling
since the parameters\break of~(\ref{arma}) can be estimated by standard
$\operatorname{ARMA}$
methods. The implication is that while the catch-all approach has some useful
robustness properties, there is no reason to expect the estimator to be efficient.

\section{Measurement Error and Nonparametric Identification}

Xia and Tong emphasize that measurement error is empirically relevant and time
series methods should take it seriously. While I agree, we also need to
acknowledge that measurement error raises many troubling problems. Of primary
importance, I believe, is the vexing issue of nonparametric identifica\-tion---
whether the parameters of interest are unique\-ly determined by the distribution
of the observables. As is known from the random sampling context, measurement
error complicates identification. In general, additional information or
structure is required to identify the parameters of an unobserved latent process.
It is not sufficient to simply introduce a new estimator.

We can see this quite simply by examining the spectral density. Suppose as
above that $x_{t}$ is the process of interest and the observed process is
$y_{t}=x_{t}+\eta_{t}$ where $\eta_{t}$ is i.i.d. measurement error with variance
$\sigma_{\eta}^{2}$. Letting $f_{x}(\lambda)$ and $f_{y}(\lambda)$ be the
spectral densities of $x_{t}$ and $y_{t}$, we know that
\[
f_{y}(\lambda)=f_{x}(\lambda)+\sigma_{\eta}^{2}.%
\]
The distribution of the observables $y_{t}$ identifies $f_{y}(\lambda)$, but
$f_{x}(\lambda)$ is not identified from knowledge of $f_{y}(\lambda)$ alone.
Under the realistic assumption that $\sigma_{\eta}^{2}$ is unknown,
$f_{x}(\lambda)$ can only be identified by knowledge of the structure of
$x_{t}$ [e.g., by knowing that $x_{t}$ is an $\operatorname{AR}(1)$ as in the example
of the previous section]. But if we acknowledge that our models for $x_{t}$
are misspecified, we should view the true $f_{x}(\lambda)$ as nonparametric
and hence without structure. It follows that the spectral density
$f_{x}(\lambda)$ is not nonparametrically identified, and thus neither is the
autocorrelation structure of $x_{t}.$

Nevertheless, some features are identified. While the spectral density is
not point identified, it is interval identified. Let $\overline{f}%
=\min_{\lambda}f_{y}(\lambda)$. Observe that
\[
f_{y}(\lambda)-\overline{f}\leq f_{x}(\lambda)\leq f_{y}(\lambda).
\]
The two bounds are identified from $f_{y}(\lambda)$, so the spectral density
$f_{x}(\lambda)$ of $x_{t}$ can be bounded within this interval. The width of
the interval is $\overline{f}=\break\min_{\lambda}f_{x}(\lambda)+\sigma_{\eta}^{2}$,
which is thus an upper bound for the measurement error variance $\sigma_{\eta
}^{2}$.

What is particularly interesting is that while the level of $f_{x}(\lambda)$
is not identified, many of its most important features are identified, specifically, the peaks and troughs. What this means is that while full
knowledge of the $x_{t}$ process is not possible, important features can be
identified from the distribution of the observables~$y_{t}.$ Knowledge of
which features are identified in the presence of measurement error and/or
misspecification helps focus attention on what can be learned about unobserved
processes from observational data.

\section*{Acknowledgments}
Research supported by the National Science Foundation.

\end{document}